\documentclass[prb,aps,twocolumn,superscriptaddress,longbibliography]{revtex4-2}

\usepackage[dvipsnames, svgnames, x11names]{xcolor}
\usepackage{amsmath}
\usepackage{amssymb}
\usepackage{graphicx}
\usepackage[colorlinks,citecolor=blue,linkcolor=blue,urlcolor=blue]{hyperref}
\usepackage{float}

\def\avg#1{\left\langle#1\right\rangle}

\begin{document}
\title{Magnetic fluctuations near the Van Hove singularity in the kagome-lattice Hubbard model at finite doping}

\author{Jingyao Wang}
\thanks{These authors contributed equally to this work.}
\affiliation{Key Laboratory for Microstructural Material Physics of Hebei Province, School of Science, Yanshan University, Qinhuangdao 066004, China}

\author{Zixuan Jia}
\thanks{These authors contributed equally to this work.}
\affiliation{School of Physics and Astronomy, and Key Laboratory of Multiscale Spin Physics (Ministry of Education), Beijing Normal University, Beijing 100875, China}

\author{Zenghui Fan}
\affiliation{School of Physics and Astronomy, and Key Laboratory of Multiscale Spin Physics (Ministry of Education), Beijing Normal University, Beijing 100875, China}

\author{Qingzhuo Duan}
\affiliation{School of Physics and Astronomy, and Key Laboratory of Multiscale Spin Physics (Ministry of Education), Beijing Normal University, Beijing 100875, China}

\author{Tianxing Ma}
\email{txma@bnu.edu.cn}
\affiliation{School of Physics and Astronomy, and Key Laboratory of Multiscale Spin Physics (Ministry of Education), Beijing Normal University, Beijing 100875, China}

\begin{abstract}
The kagome-lattice Hubbard model attracts widespread interest due to its flat-band and Van Hove singularity features, which can give rise to unconventional magnetism. We employ determinant quantum Monte Carlo simulations to systematically investigate the uniform magnetic susceptibility across a range of on-site interactions and electron fillings on a two-dimensional kagome lattice. 
Beyond the Van Hove singularity, dominant ferromagnetic fluctuations emerge.
Magnetic susceptibility grows markedly with increasing interaction strength and decreasing temperature, indicating that the Van Hove singularity acts as a critical point for 
the crossover of dominant magnetic fluctuations. 
Finite-size analysis further suggests the potential stabilization of a finite-temperature ferromagnetic phase. We also examine the sign problem to identify numerically reliable parameter regimes. 
These results provide valuable insights into controlling magnetic fluctuations in kagome systems and establish a computational framework for exploring flat-band physics in regimes characterized by novel quantum phases and competing orders.
\end{abstract}
\maketitle 

\section{Introduction}

In recent years, kagome lattice systems such as $A\mathrm{V}_3\mathrm{Sb}_5$ \cite{PhysRevLett.125.247002,Chen2021,Neupert2022,PhysRevLett.127.046401,PhysRevB.105.L100502} and CoSn-type materials  \cite{PhysRevB.102.075148,PhysRevLett.128.096601} have attracted extensive research interest due to their correlated electronic phenomena, including superconductivity \cite{PhysRevLett.125.247002}, charge density waves \cite{Luo2022}, and the anomalous Hall effect \cite{doi:10.1126/sciadv.abb6003}. 
The unique lattice geometry and strong electronic correlations in these kagome materials make them suitable platforms for investigating the competition and evolution of quantum states. 
Theoretical studies predict that kagome lattices can host flat bands \cite{PhysRevLett.126.196403}, nontrivial topological states \cite{Jiang2021}, and time-reversal symmetry-breaking charge orders \cite{Mielke2022}. Experimental investigations have further confirmed the existence of macroscopic quantum orders such as pair density waves \cite{Chen2021} and nematicity \cite{Nie2022}. 
The physics of kagome lattices originates from both strong electronic correlations and their intrinsic structure, which simultaneously generates Dirac cones, flat topological bands, and Van Hove singularity (VHS)\cite{10.1143/ptp/6.3.306,Kang2020,PhysRevLett.121.096401,Yin2019,PhysRevMaterials.3.094407,10.1093/nsr/nwac199,PhysRevB.108.235163}. 
The interplay of these features collectively produces nontrivial electronic topology and correlated many-body quantum states.
The magnetic properties of kagome lattice systems constitute another frontier of fundamental interest. In quantum magnets, the kagome lattice gives rise to extraordinary magnetic phenomena \cite{Ramirez2003}. Particularly in kagome compounds\cite{PhysRevLett.121.096401,PhysRevX.11.031050,PhysRevLett.127.177001,PhysRevLett.127.217601}, this magnetism manifests as exotic quantum ground states, notably quantum spin liquids\cite{PhysRevB.110.085146} and spin ice states \cite{Balents2010,PhysRevB.67.235102,PhysRevLett.89.226402}.
Remarkably, the characteristic flat bands of kagome metals offers the possibility of realizing a strong ferromagnetic correlation \cite{AMielke_1991,PhysRevLett.100.136404}.

\begin{figure}[tp]
	\includegraphics[width=\columnwidth]{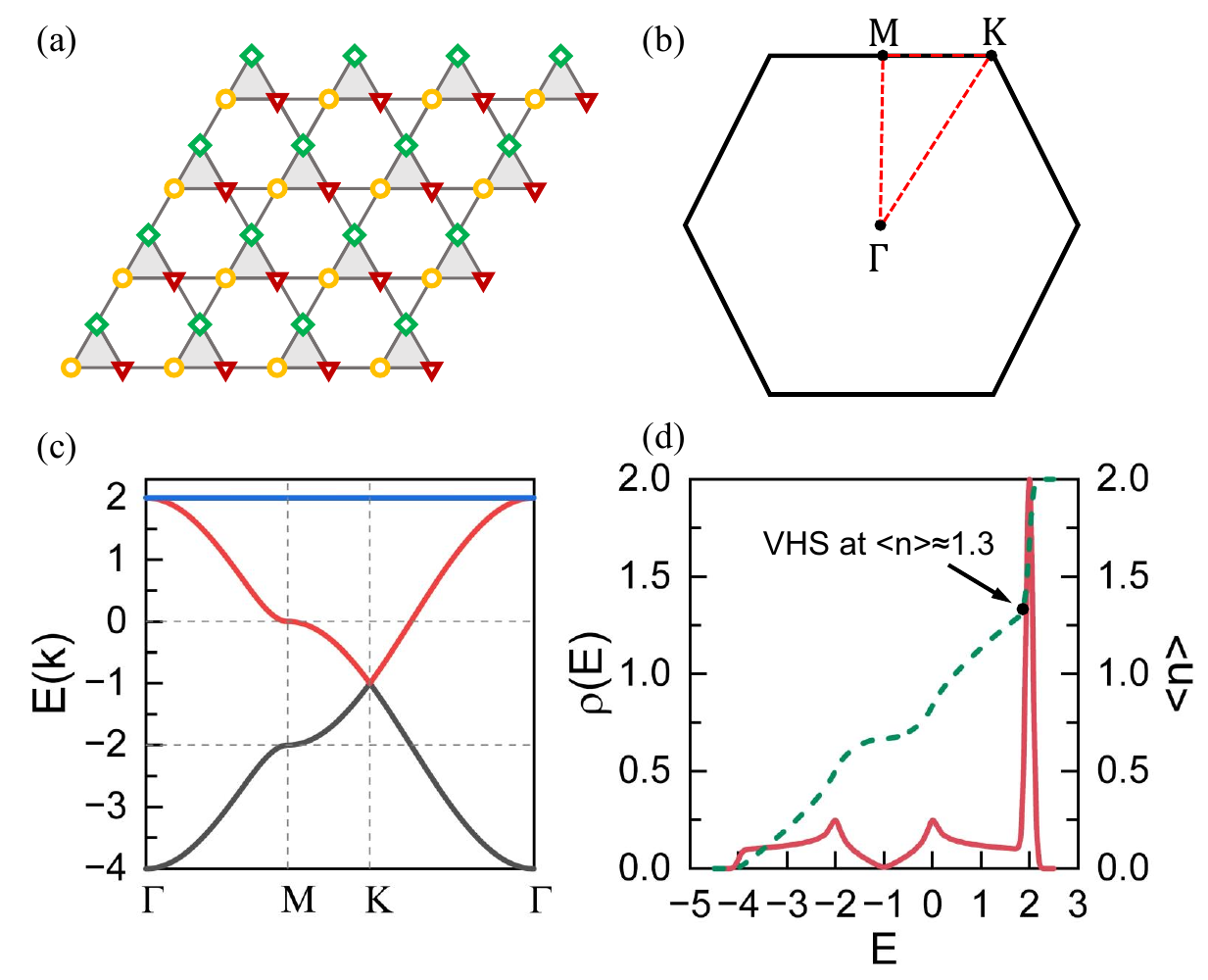}
	\caption{(a) The structure of the kagome lattice. In each unit cell (gray triangles), the sublattice is labeled as A (yellow dots), B (red dots), and C (green dots). (b) The Brillouin zone of the kagome lattice. (c) The band structure of the tight-binding kagome model along the high-symmetry direction. (d) Density of states (DOS) (solid lines) and filling $\langle n\rangle$ (dashed lines) as functions of energy. }
	\label{fig:lattice}
\end{figure}

	Flat-band systems have attracted substantial interest because strong correlations can lead to unconventional quantum phases when the kinetic energy is strongly suppressed. 
	In moir\'e flat-band systems, large-scale quantum Monte Carlo (QMC) simulations have revealed superconductivity and a high-temperature transition from a bosonic fluid to a pseudogap\cite{PhysRevB.106.184517}.
	This naturally raises the question of how other flat-band platforms show correlation effects beyond the moir\'e setting, especially in magnetism.
The intrinsic bands of kagome lattices simultaneously host a flat band, two VHSs at the $M$ point (Fig.\ref{fig:lattice}), and geometric frustration\cite{PhysRevLett.121.096401,PhysRevX.11.031050}. The coexistence of the flat band and the VHS enhances the density of states and may strongly affect magnetic correlations. Experimentally, kagome materials such as CoSn and $A\mathrm{V}_3\mathrm{Sb}_5$ have been observed to exhibit charge density waves and anomalous Hall effects driven by VHS\cite{Luo2022, doi:10.1126/sciadv.abb6003}. These facts motivate a systematic numerical study of whether the VHS sets a crossover scale for magnetic correlations and whether it promotes ferromagnetic fluctuations in the kagome Hubbard model.

Recent advances in magnetic order manipulation through charge transfer have sparked growing interest in two-dimensional material systems. Studies on doped graphene-based materials\cite{Cervenka2009} and doped transition metal dichalcogenides \cite{Reyntjens_2021,PhysRevB.107.245126} have demonstrated effective magnetic order control, establishing doped semiconductor two-dimensional (2D) materials as promising candidates for next-generation spintronic applications and sensing technologies.
In the doped kagome Hubbard model, the cooperative regulation of electron filling concentration and Coulomb interaction provides a controllable platform for exploring the evolution of quantum magnetism. Recent studies on doped kagome systems highlight profound doping-dependent magnetic evolution. 
Density matrixrenormalization group (DMRG) results show light doping stabilizes insulating charge density wave states with short-range spin-spin correlations in kagome quantum spin liquids\cite{https://doi.org/10.1002/qute.202000126}. 
Fermionic projected entangled simplex state (PESS) simulations reveal hole doping induces transitions from charge order to emergent magnetic phases via altered magnetic correlations\cite{xu2024globalphasediagramdoped}. 
Dynamical mean-field theory (DMFT) studies establish doping and interaction-driven competition between spiral magnetic orders\cite{PhysRevLett.134.066502}.
Experimentally, magnetic responses exhibit tunable correlation strength. 
Hole doping in CoSn enhances magnetic correlations and induces spin-glass states through flat-band shifts \cite{PhysRevMaterials.5.044202}, 
while LaCuO$_{2.66}$ manifests doping-induced inhomogeneous magnetism with coexisting nonmagnetic sites and frozen moments \cite{PhysRevB.87.214423}. 
These findings motivate our investigation into the emergence of dominant magnetic fluctuations in the doped kagome Hubbard model, presenting a compelling research direction.

In the kagome lattice, there are three nonequivalent sites per unit cell of the square lattice, defining three interpenetrating sublattices, as illustrated in Fig.\ref{fig:lattice}(a)
Figure \ref{fig:lattice}(b) displays the Brillouin zone of the kagome lattice, where $K$, $M$, and $\Gamma$ represent the high-symmetry points in the first Brillouin zone. 
As seen in Fig.\ref{fig:lattice}(c), the kagome lattice features three bands, including two cone-shaped Dirac bands and one distinctive flat band. At the $K$ point, there is a Dirac point, while at $M$ point, there are two VHSs.
The density of states is shown in Fig.\ref{fig:lattice}(d). It shows that there exists a VHS at $\langle n\rangle$$\approx$1.3. Usually, the interplay between electronic correlation and flat band, or VHS, may lead to strong ferromagnetic fluctuations. 
The magnetic exchange between local spins is largely dominated by the contribution of the flat band, which is ferromagnetic, as revealed by many studies \cite{PhysRevB.103.165105,PhysRevB.85.205104,LIU20191490,PhysRevB.104.155129,PhysRevB.106.155142}.

In this work, we present comprehensive numerical simulations of magnetic correlation in the doped Hubbard model on a kagome lattice using the determinant quantum Monte Carlo (DQMC) method. 
The kagome lattice configuration we consider can be realized and tuned in ultracold-atom experiments. 
Specifically, Jo \textit{et al.} constructed a two-dimensional kagome lattice by superposing two commensurate triangular lattices and tuning their relative displacement \cite{PhysRevLett.108.045305}. In such setups, the key Hubbard parameters are experimentally controllable.  
Our finite-temperature results can therefore be directly compared with measurements of spin correlations and susceptibilities, providing guidance for exploring doping-driven magnetic correlations in kagome optical lattices.
We also address the sign problem to identify which parameter regimes yield reliable results. 
	We focus on the relation between the VHS and magnetic fluctuations in the kagome Hubbard model. We test whether the VHS filling $\langle n\rangle \approx 1.3$ marks a crossover in magnetic correlations, and whether the intrinsic flat band together with the VHS strengthens ferromagnetic tendencies. Within accessible temperatures and system sizes, we also examine whether fillings beyond the VHS show a stronger tendency toward finite-temperature ferromagnetism. 
Notably, we find that the system exhibits strong ferromagnetic fluctuations once the electron filling reaches VHS $\langle n\rangle$$\ge$1.3. Finite-size scaling analysis suggests the potential existence of a finite-temperature ferromagnetic phase. The pronounced ferromagnetic fluctuations we observe can be understood within the flat-band framework.
In this way, our results add a magnetic viewpoint to flat-band physics in an intrinsic kagome platform.

\section{MODEL AND METHODS}
We consider the Hubbard model defined on a 2D kagome lattice. The kagome lattice possesses three sublattices, so the number of total lattice sites is $3\times L^2$. The Hamiltonian reads 
\begin{equation}
H=-t\sum_{\langle i,j\rangle\sigma}(\hat{c}_{i\sigma}^{\dagger}\hat{c}_{j\sigma} + \text{H.c.})+U\sum_{i}\hat{n}_{i\uparrow}\hat{n}_{i\downarrow}
-\mu\sum_{i\sigma}\hat{n}_{i\sigma}.
\end{equation} 
where $t$ is the hopping amplitude between the nearest neighbor (NN) sites on the lattice. We use $\langle i,j\rangle$ to denote the nearest-neighbor sites. To simplify calculations, we set $t$=1, which serves as the unit of energy. $i$ and $j$ index the lattice sites in the full $3\times L^2$ sites. $\hat{c}_{i\sigma}^{\dagger}(\hat{c}_{i\sigma})$ creates (annihilates) electrons at site $i$ in the kagome lattice with spin $\sigma$ $(\sigma=\uparrow,\downarrow)$. $\hat{n}_{i\sigma}=\hat{c}_{i\sigma}^\dagger \hat{c}_{i\sigma}$ denotes the electron number of spin $\sigma$ at site $i$. $U$ is the on-site Hubbard interaction. The chemical potential $\mu$ regulates the electron filling $\langle n\rangle$ of the system. $\langle n\rangle$ represents the average electron number per site.

 	The determinant quantum Monte Carlo method\cite{santos_2003,PhysRevB.28.4059,PhysRevD.24.2278,PhysRevD.24.2278} probes the finite-temperature properties of the Hubbard model on a kagome lattice. The inverse temperature $\beta$ is discretized into $L_\tau$ imaginary-time slices via Trotter-Suzuki decomposition, approximating the partition function
 	$ Z = \operatorname{Tr}\left[e^{-\beta H}\right] $
 	as a product of non commuting exponentials. The Hubbard-Stratonovich (HS) transformation decouples quartic electron-electron interactions via auxiliary scalar fields $s_{i}(l)$, reducing the Hamiltonian to a fermionic quadratic form.
 	The weight of an HS field configuration scales with the product of spin-up/spin-down fermion determinants: $\mathbf{P}_s = \det\left[I + B_s(\beta, 0)\right] / {\sum_{s} \det(I+B_{s}(\beta, 0))}$. Monte Carlo sampling employs the Metropolis algorithm, governed by the acceptance ratio $R = \mathbf{P}_{s'}/\mathbf{P}_s$. Observables are evaluated via the single-particle Green’s function $G^{\sigma} = \left[I + \prod_{l} B_{l, \sigma} \right]^{-1}$, where $B_{l,\sigma} = e^{t \Delta \tau K} e^{\sigma \nu V_{l}}$ combines the non interacting Hamiltonian $K$ and HS field coupling $V_l$.
 	Statistical uncertainties are mitigated by averaging over $2\times10^5$ measurements\cite{PhysRevB.85.125127}. Systematic Trotter errors [$\propto(\Delta\tau)^2$] are constrained via a sufficiently small $\Delta\tau$ (consistent with Ref. \cite{PhysRevB.85.125127}). The DQMC method is well validated for systems including doped graphene\cite{10.1063/1.3485059}, iron-based superconductors\cite{PhysRevLett.110.107002}, and geometrically frustrated lattices\cite{PhysRevB.80.014428}.

\begin{figure}[tp]
	\centering
	\includegraphics[width=0.9\columnwidth]{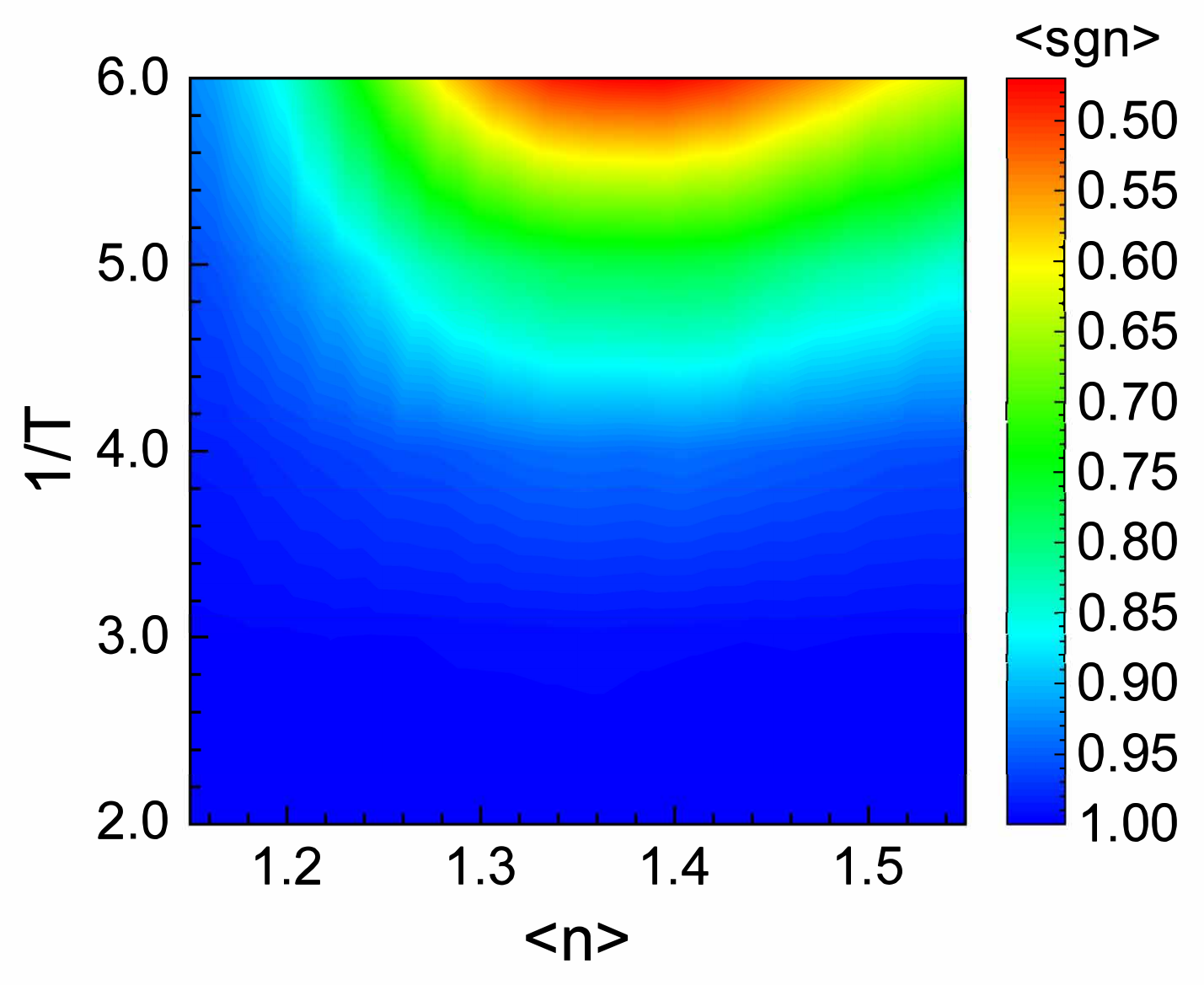}
	\caption{The average sign as a function of electron filling $\avg{n}$ for different temperature $T$. Even at the lowest temperatures, ⟨sign⟩ remains above 0.45 in the crossover regime, ensuring numerical reliability. The error is controlled within 4.4$\times$10$^{-4}$.}
	\label{fig:sign}
\end{figure}

To investigate magnetic ordering tendencies in the system, we analyze the zero-frequency spin susceptibility along the $z$ axis defined by
\begin{equation}
\chi\left(q\right)=\frac{1}{N}\sum_{i,j}\sum_{l,m}\int_0^\beta d{\tau} e^{-iq\left(\mathbf{R}_i-\mathbf{R}_j\right)}\left\langle S_{i,l}^z\left(\tau\right)S_{j,m}^z\left(0\right)\right\rangle
\end{equation}
where the time-dependent spin operator $S_{i,l}^{z}(\tau)=e^{H\tau}S_{j,m}^{z}(0)e^{-H\tau}$ represents the $z$-component of electron spin with
$S_{i,l}^{z}=c_{il\uparrow}^{\dagger}c_{il\uparrow}-c_{il\downarrow}^{\dagger}c_{il\downarrow}$. 
The calculations are performed for lattice dimensions from $L=4$ to $L=8$ to assess finite-size effects.

We calculate the average sign using the following expression\cite{PhysRevB.92.045110} to address the sign problem, 
\begin{equation}
\langle \text{sgn} \rangle = \frac{\sum_{\mathcal{X}} \det M_\uparrow (\mathcal{X}) \det M_\downarrow (\mathcal{X})}{\sum_{\mathcal{X}} |\det M_\uparrow (\mathcal{X}) \det M_\downarrow (\mathcal{X})|}
\end{equation}
where $M_\sigma (\mathcal{X})$ represents each spin species matrix. When ⟨sign⟩ approaches 1, it indicates a milder sign problem and higher reliability of the numerical results.

\begin{figure}[tp]
\centering
\includegraphics[width=0.9\columnwidth]{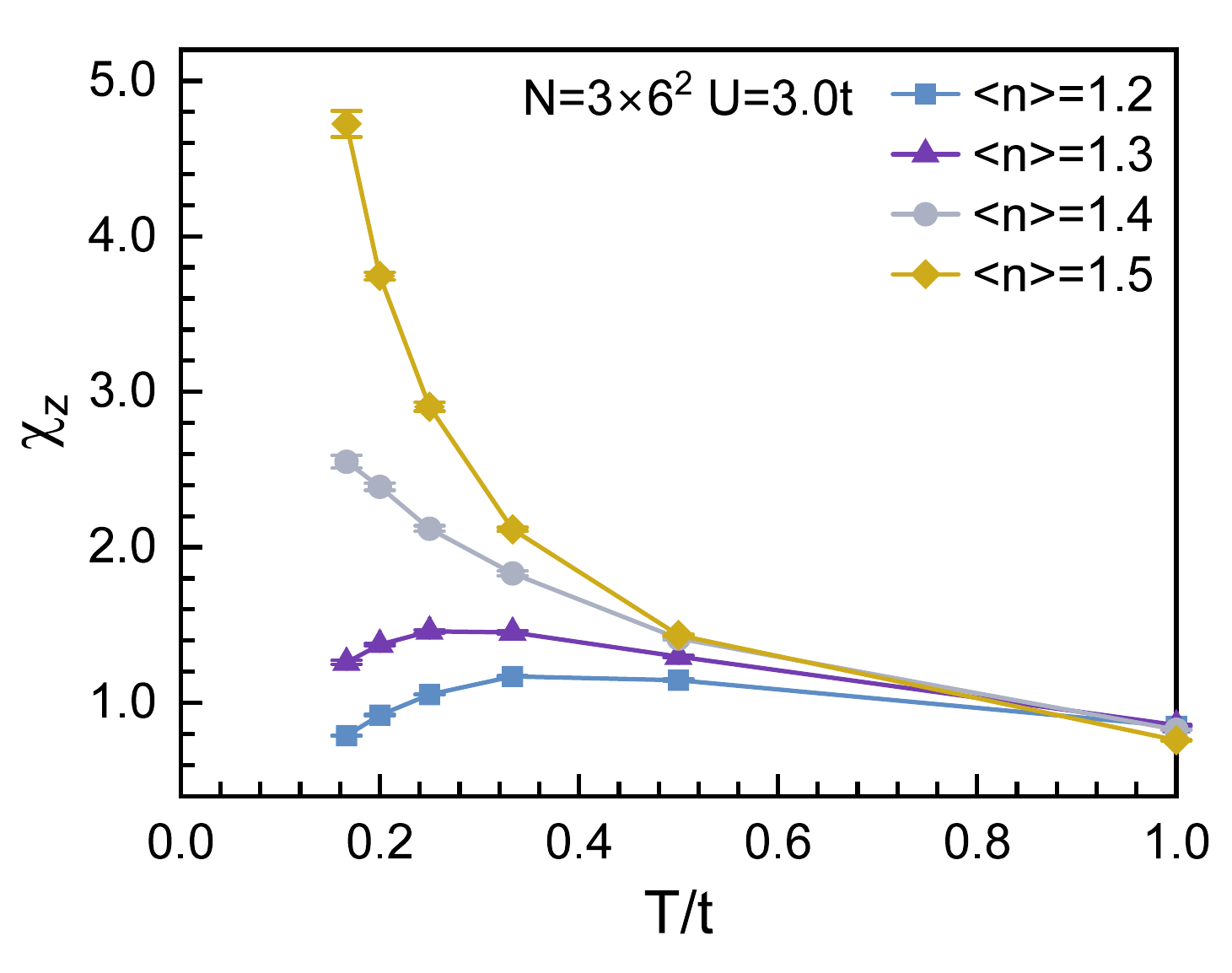}
\caption{The temperature-dependent z-component spin susceptibility $\chi_z$ at $U$=3.0$t$ with different fillings $\avg{n}$.}
\label{fig:3}
\end{figure}

\begin{figure}[b]
	\centering
	\includegraphics[width=0.9\columnwidth]{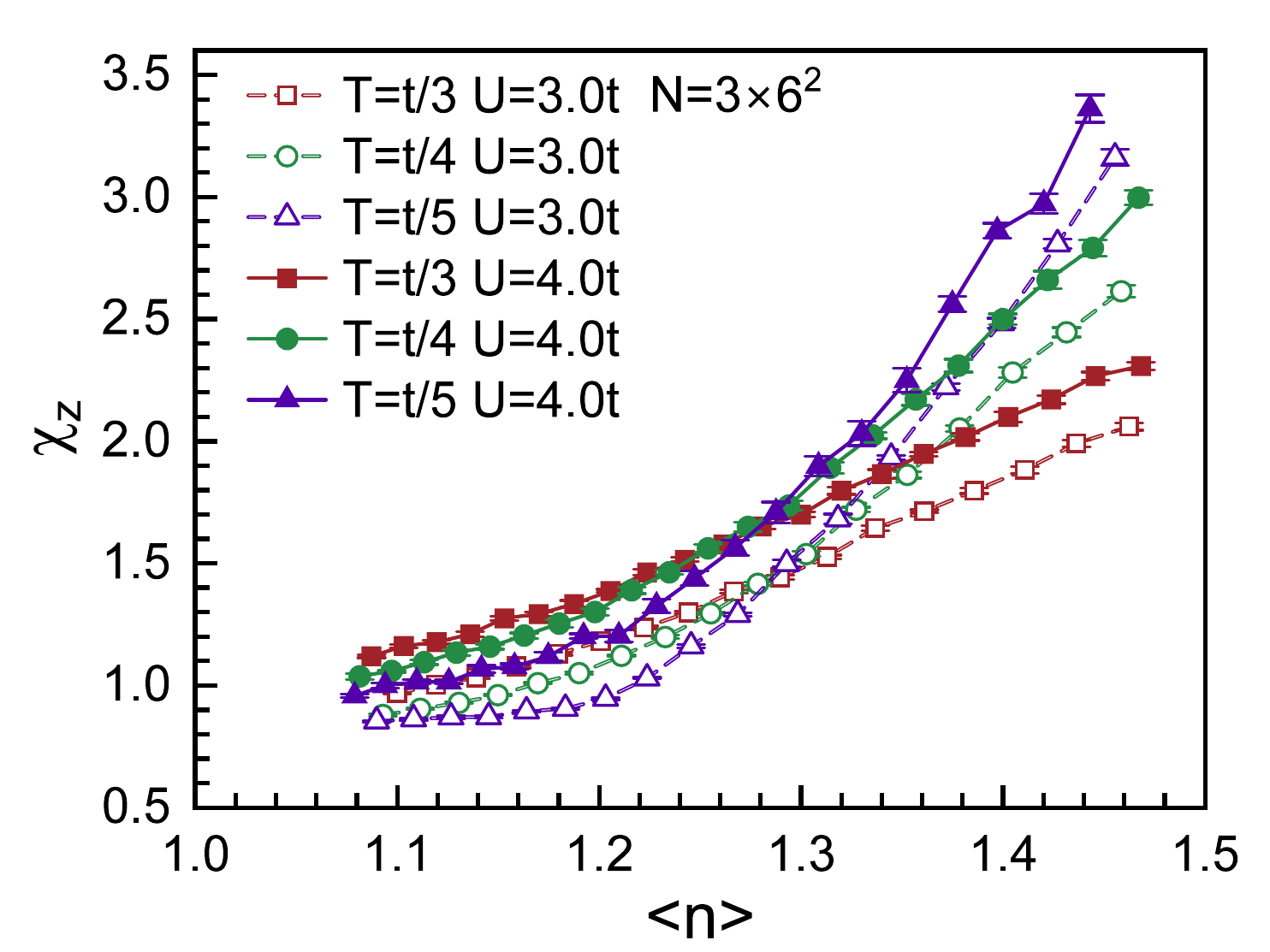}
	\caption{The uniform spin magnetic susceptibility vs electron doping concentration at different values of $T$ with $U=3.0t$ and $U=4.0t$. }
	\label{fig:4}
\end{figure}

	According to previous studies, the sign problem critically limits QMC simulations at low temperatures and in strong-coupling regimes. However, this can probably be circumvented by calculating much longer runs. 
Figure \ref{fig:sign} evaluates the average sign, $\avg{\text{sgn}}$ across parameter space to identify reliable simulation areas unaffected by the sign problem. 
The data reveal non monotonic evolution of $\avg{\text{sgn}}$ with decreasing temperature and increasing doping concentration $\avg{n}$, exhibiting an obvious minimum within the critical range 1.3$<$$\avg{n}$$<$1.4 where sign becomes relatively severe.
Remarkably, this minimum coincides precisely with the doping regime dominated by ferromagnetic fluctuations. Spin susceptibility analysis in the following Fig. \ref{fig:4} independently confirms this behavior for $\avg{n}$$>$1.30.
Given the reported correlation between $\avg{\text{sgn}}$ suppression and quantum phase transitions \cite{doi:10.1126/science.abg9299},
the observed minimum likely originates from enhanced quantum critical fluctuations which occur proximate to ferromagnetic ordering.
We implemented expanded measurement protocols up to 10000–200000 iterations in severe sign problem conditions. This ensures statistical reliability. The protocols mitigate fluctuations effectively and guarantee data accuracy\cite{HUANG2019310}.

\section{RESULTS AND DISCUSSION}   

As shown in Fig. \ref{fig:3}, when the filling $\avg{n}$$\le$1.3, the spin susceptibility $\chi_z$=$\chi(\Gamma)$ levels off as temperature decreases, indicating negligible magnetic tendencies, where $\chi_z$ denotes the uniform spin susceptibility at the $\Gamma$ point in the Brillouin zone; by contrast, for $\avg{n}$$\ge$1.4,  $\chi_z$  increases steadily as temperature drops, signaling pronounced ferromagnetic correlations. This clear contrast near the VHS suggests that ferromagnetic ordering emerges once the filling enters the critical range 1.3$\le$$\avg{n}$$\le$1.4. Moreover, raising the interaction from $U=3.0t$ to $4.0t$ further enhances $\chi_z$, and in both cases, magnetic susceptibility curves at different temperatures intersect in the vicinity of $\avg{n}$$\approx$1.3, highlighting the VHS point as a magnetic crossover.

To understand the filling dependence of magnetic correlations in a kagome lattice, we show the uniform spin susceptibilities $\chi_z$ with different values of $T$ in Fig. \ref{fig:4}.
The uniform spin susceptibility exhibits a temperature-dependent enhancement with increasing electron doping $\avg{n}$, displaying accelerated growth kinetics at lower temperatures that follows a quasi-exponential scaling. Strikingly, all temperature-dependent curves converge at a critical electron filling concentration $\avg{n}$$\approx$1.3 under $U=3.0t$, where the spin susceptibility becomes temperature independent. This universal crossing point marks the crossover regime between paramagnetic and emergent ferromagnetic ordering. 
Notably, the critical filling $\avg{n}$$\approx$1.3 coincides with the interval 1.3$<$$\avg{n}$$<$1.4, where the average sign $\avg{\text{sgn}}$ exhibits a clear minimum. 
It is argued that the sign problem in determinant quantum Monte Carlo is quantitatively linked to quantum critical behavior\cite{doi:10.1126/science.abg9299}. Therefore, the observed coincidence suggests that the reduction of $\langle \text{sgn} \rangle$ at this doping may be tied to enhanced ferromagnetic fluctuations. This further supports the possibility that quantum critical fluctuations near ferromagnetic ordering could play a governing role in the behavior of the sign problem.
Crucially, the critical $\avg{n}$ coincides with the electron filling corresponding to the VHS, establishing direct correlation between electronic topology and magnetic phase transition.
At $U$=4.0$t$, the spin susceptibility follows similar evolution patterns with electron doping and temperature as before, with spin susceptibility curves at different temperatures still intersecting near $\avg{n}$$\approx$1.3. Due to enhanced strong correlation effects, the overall magnitude of spin susceptibility increases compared to the $U=3.0t$ case. This demonstrates that strengthening Coulomb interaction enhances spin susceptibility,
which may further amplify the quantum critical fluctuations associated with the sign problem in this doping regime. 

\begin{figure}[t]
\centering
\includegraphics[width=0.9\columnwidth]{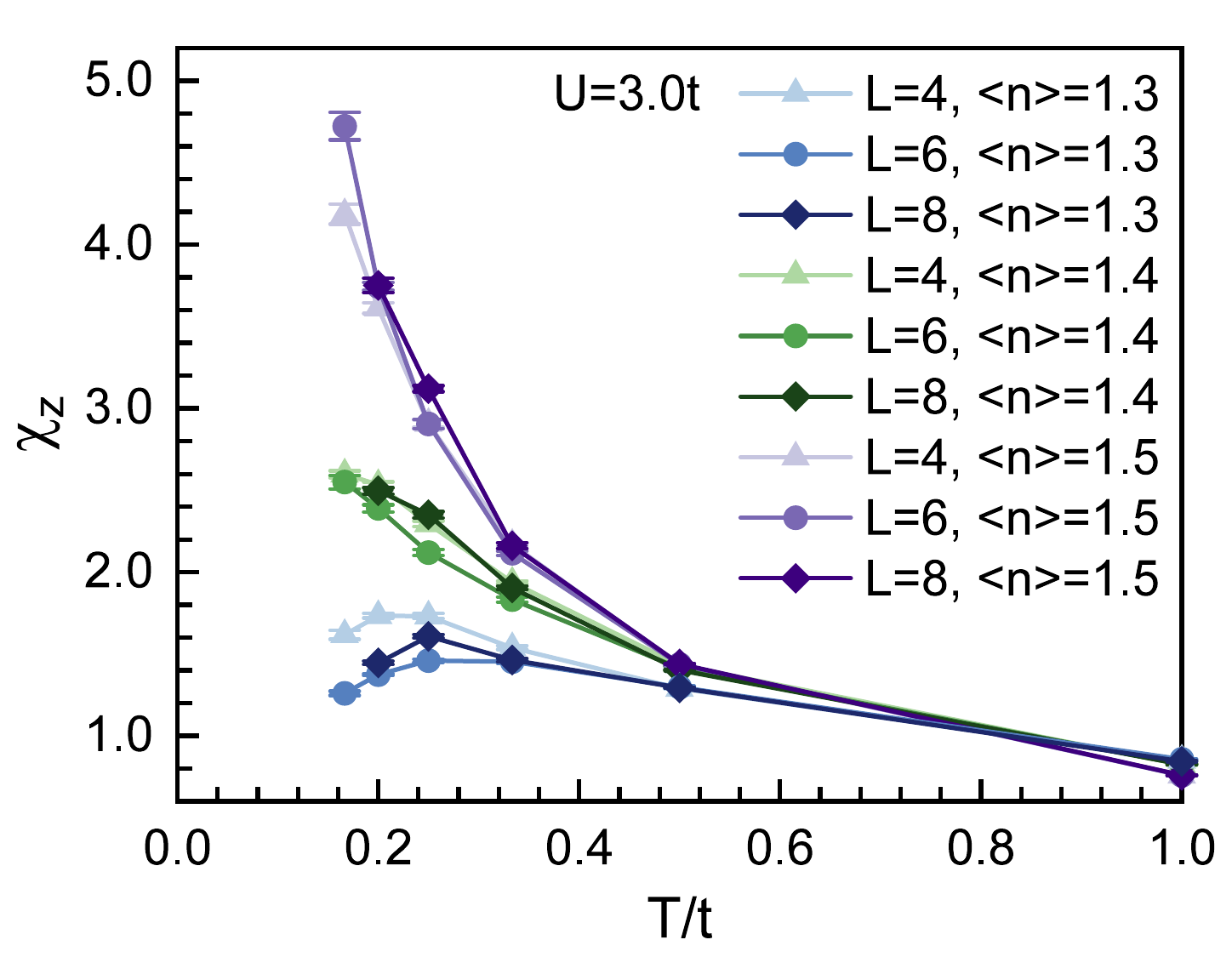}
\caption{Uniform spin susceptibility $\chi_z$ as a function of $T$ with various $L=4,6,8$ at $\avg{n}=1.3, 1.4, 1.5$. Calculations are based on $U=3.0t$ system.}
\label{fig:5}
\end{figure}

\begin{figure}[b]
	\centering
	\includegraphics[width=0.95\columnwidth]{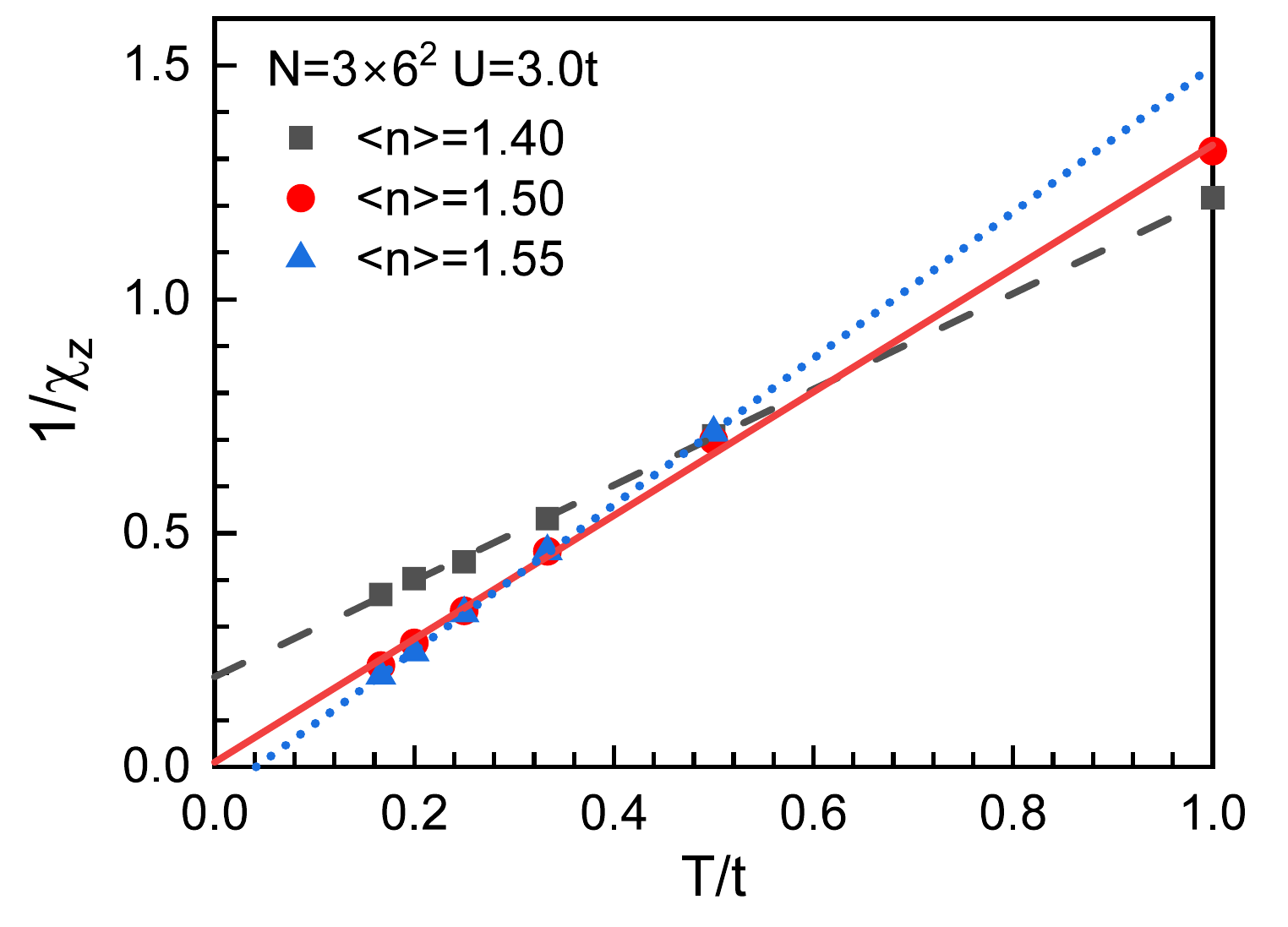}
	\caption{The temperature-dependent $1/\chi_z$ at $U=3.0t$ with different $\avg{n}$.  A finite $T_C$ for the possible transition into the ferromagnetic phase requires that the extrapolated line intersect the $y$ axis at a negative value. } 
	\label{fig:6}
\end{figure}

\begin{figure*}[t]
	\centering
	\includegraphics[width=\textwidth]{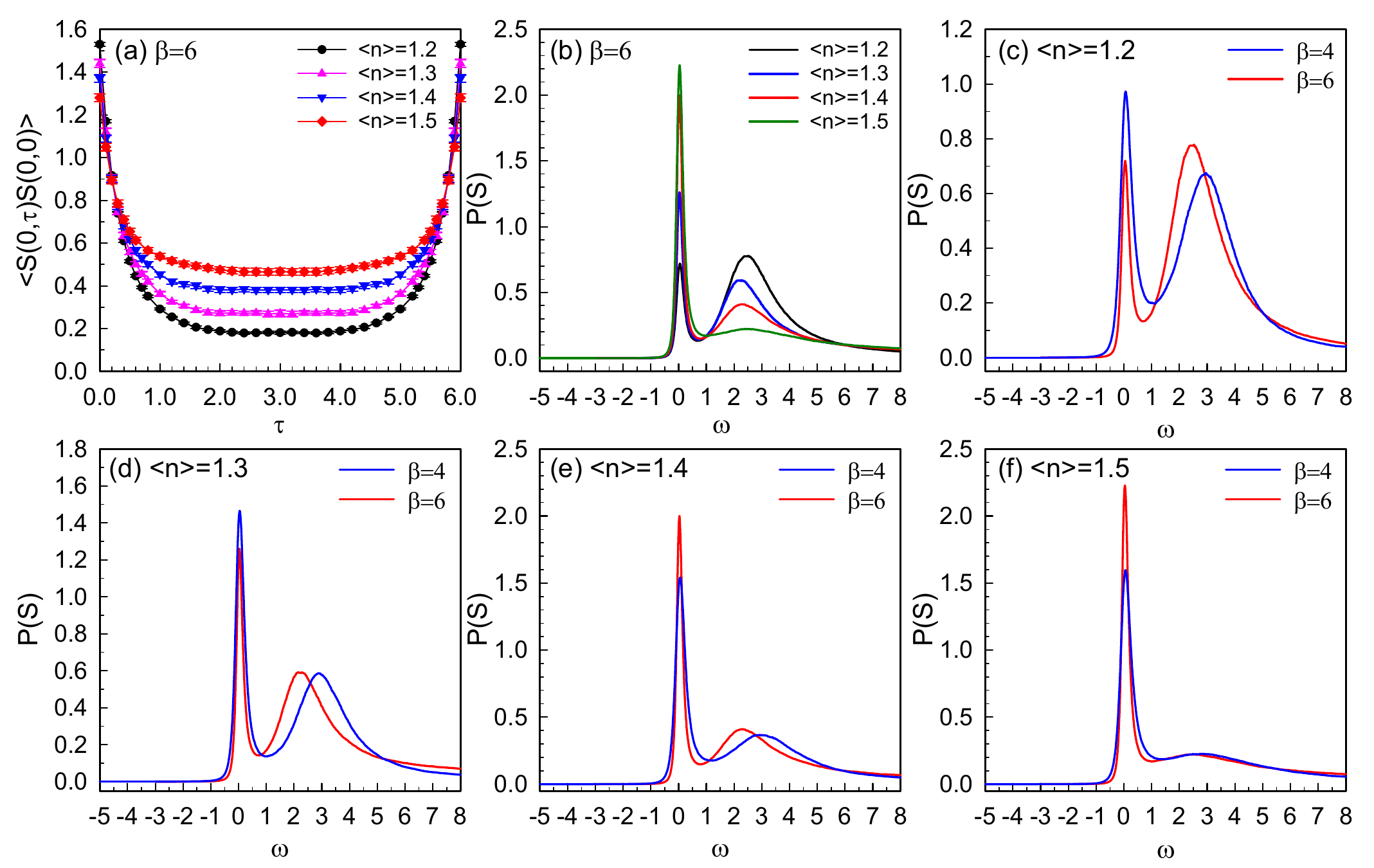}
	\caption{The time-dependent spin correlation functions are shown in (a).  (b) plots the spectra at electron densities \( n = 1.2, 1.3, 1.4, 1.5 \) and $\beta=6$.   (c)-(f) plot the spectra for $\beta=4$ and 6 at different electron densities, respectively. }
	\label{fig:7}
\end{figure*}
As shown in Fig. \ref{fig:5}, we examine the uniform magnetic spin susceptibility as a function of temperature $T$ across different lattice sizes to assess finite-size effects on our physical findings. Figure \ref{fig:5} displays the temperature dependence of uniform spin susceptibility $\raisebox{0.5ex}{$\chi$}_z$.  
Near the VHS, the $\raisebox{0.5ex}{$\chi$}_z$ curves for all the lattice sizes exhibit minor numerical variations but uniformly converge as temperature decreases. Upon increasing the doped electron filling to $\langle n \rangle=1.4$ and 1.5, the magnetic susceptibility shows a consistent divergent behavior for all system sizes, further confirming that finite-size effects are negligible.

In Fig. \ref{fig:6}, we present the temperature dependence of uniform spin susceptibility at $U=3.0t$ for different electron fillings $\avg{n}$. 
The data exhibit a linear correlation between $1/\chi_z$ and temperature $T$, consistent with Curie-Weiss behavior $1/\chi_z = (T - T_C)/A$.
We thus extrapolate $1/\chi_z$ to zero temperature via linear fitting. This corresponds to infinite uniform spin susceptibility $\chi$, defining the critical temperature $T_C$. For $\avg{n}=1.40$, the extrapolated intercept remains positive, indicating no appreciable ferromagnetic transition has developed.
Systematically, the $y$-intercept shifts downward with increasing $\avg{n}$.
Negative intercepts emerge at $\avg{n}$$\approx$1.50.
A definitive negative $y$ intercept appears at $\avg{n}=1.55$.
This progression of negative intercepts signifies intersections with the temperature axis, implying potential spontaneous ferromagnetic ordering at finite temperatures. Crucially, a finite $T_C$ requires a negative $y$ intercept in this framework.
Figure \ref{fig:6} suggests enhanced potential for $T_C$ formation at elevated $\avg{n}$, where the electron concentration surpasses the VHS position.  

In Fig. 7, we present the time-dependent spin correlation functions and the corresponding spectra at inverse temperatures $\beta=4, 6$, for electron densities $n=1.2, 1.3, 1.4, 1.5$.
	The flat band hosts strong ferromagnetic fluctuations. For $n=1.2$ and $n=1.3$, the Fermi level ($\omega=0$) is still below the flat band. In this regime, only a small fraction of electrons is thermally excited into the flat band. Because the flat band is broadened by thermal fluctuations and the on-site interaction $U$, these excitations generate a ferromagnetic peak at finite (relatively high) energy.
	For $n=1.4$ and $n=1.5$, the Fermi level enters the flat band. As a result, the finite-energy ferromagnetic peak is gradually suppressed. At the same time, the spectral weight near $\omega=0$ becomes dominant, indicating pronounced ferromagnetic fluctuations inside the flat band. 

\section{Summary} 
In this work, we investigate magnetic behavior in a doped kagome-lattice Hubbard model using the DQMC method and present exact numerical results on the magnetic correlations. We explore how electron filling and interaction strength affect spin fluctuations near the VHS and whether ferromagnetic correlations emerge. Our key finding is that doping beyond the VHS induces ferromagnetic spin fluctuations, identifying the VHS as a quantum critical point separating paramagnetic and ferromagnetic-fluctuating regimes. We verify the robustness of these results against lattice size. A Curie-Weiss analysis at \(U=3.0t\) yields positive extrapolated Curie temperatures for \(\langle n\rangle \ge 1.5\), suggesting the onset of a finite-temperature ferromagnetic phase. We also analyze the sign problem and describe the parameter regimes where our simulations are controlled, and show that reliable information can still be extracted in the vicinity of the critical filling.
These findings provide a microscopic reference for interpreting filling-controlled magnetism in kagome-based platforms. The magnetic response of kagome metal and magnets can change under carrier doping or pressure\cite{D5MH00120J, Wang2025}. Recent studies on titanium-based kagome metals LnTi\(_3\)Bi\(_4\) reported strong coupling between itinerant electrons and magnetism, together with unusual magneto transport\cite{Cheng2025}. In this context, our DQMC results offer a direct filling-based criterion.
Overall, our results not only offer valuable insights for exploring the unique flat-band characteristics of kagome lattices but also clarify how various magnetic orders compete in strongly correlated materials, guiding future experimental work in this regime.

\acknowledgments
This work was supported by the National Natural Science Foundation of China under Grants No. 12504182 and No. 12474218 and Beijing Natural Science Foundation (Grants No. 1242022 and No. 1252022). 
J.W. acknowledges financial support from the Natural Science Foundation of Hebei Province (Grant No. A2025203038) and Science Research Project of Hebei Education Department (Grant No. QN2025017). 

\section*{Data availability} 
The data that support the findings of this article are openly available\cite{wang_2025_15667105}.

\bibliography{BibTex_Refs}
\end{document}